\documentclass[twocolumn]{aastex63}

\graphicspath{{./}{figures/}}

\received{September 20, 2019}
\revised{November 22, 2019}
\accepted{November 27, 2019}
\submitjournal{ApJS}

\shorttitle{Inner Heliosphere Type III Bursts}
\shortauthors{Pulupa et al.}
\begin{document}

\title{Statistics and Polarization of Type III Radio Bursts Observed in the Inner Heliosphere}

\correspondingauthor{Marc Pulupa}
\email{pulupa@berkeley.edu}

\author[0000-0002-1573-7457]{Marc Pulupa}
\affil{Space Sciences Laboratory, University of California, Berkeley, CA 94720-7450, USA}

%\author[0000-0002-1989-3596]{Stuart D. Bale}
%\affil{Physics Department, University of California, Berkeley, CA 94720-7300, USA}
%\affil{Space Sciences Laboratory, University of California, Berkeley, CA 94720-7450, USA}
%\affil{The Blackett Laboratory, Imperial College London, London, SW7 2AZ, UK}

\author[0000-0002-1989-3596]{Stuart D. Bale}
\affil{Physics Department, University of California, Berkeley, CA 94720-7300, USA}
\affil{Space Sciences Laboratory, University of California, Berkeley, CA 94720-7450, USA}
\affil{The Blackett Laboratory, Imperial College London, London, SW7 2AZ, UK}
\affil{School of Physics and Astronomy, Queen Mary University of London, London E1 4NS, UK}

\author[0000-0002-6145-436X]{Samuel T. Badman}
\affil{Physics Department, University of California, Berkeley, CA 94720-7300, USA}
\affil{Space Sciences Laboratory, University of California, Berkeley, CA 94720-7450, USA}

\author[0000-0002-0675-7907]{J. W. Bonnell}
\affil{Space Sciences Laboratory, University of California, Berkeley, CA 94720-7450, USA}

\author[0000-0002-3520-4041]{Anthony W. Case}
\affil{Smithsonian Astrophysical Observatory, Cambridge, MA 02138, USA}

\author[0000-0002-4401-0943]{Thierry {Dudok de Wit}}
\affil{LPC2E, CNRS and University of Orl\'eans, Orl\'eans, France}

\author[0000-0003-0420-3633]{Keith Goetz}
\affiliation{School of Physics and Astronomy, University of Minnesota, 
Minneapolis, MN 55455, USA}

\author[0000-0002-6938-0166]{Peter R. Harvey}
\affil{Space Sciences Laboratory, University of California, Berkeley, CA 
94720-7450, USA}

\author[0000-0001-6247-6934]{Alexander M. Hegedus}
\affiliation{Climate and Space Sciences and Engineering, University of Michigan, Ann Arbor, MI 48109, USA}

\author[0000-0002-7077-930X]{Justin C. Kasper}
\affiliation{Climate and Space Sciences and Engineering, University of Michigan, Ann Arbor, MI 48109, USA}
\affiliation{Smithsonian Astrophysical Observatory, Cambridge, MA 02138 USA}

\author[0000-0001-6095-2490]{Kelly E. Korreck}
\affil{Smithsonian Astrophysical Observatory, Cambridge, MA 02138, USA}

\author[0000-0002-6809-6219]{Vladimir Krasnoselskikh}
\affil{LPC2E, CNRS and University of Orl\'eans, Orl\'eans, France}

\author[0000-0001-5030-6030]{Davin Larson}
\affil{Space Sciences Laboratory, University of California, Berkeley, CA 94720-7450, USA}

\author{Alain Lecacheux}
\affil{LESIA, Observatoire de Paris, Universit\'{e} PSL, CNRS, Sorbonne Universit\'{e}, Universit\'{e} de Paris, 5 place Jules Janssen, 92195 Meudon,France}

\author[0000-0002-0396-0547]{Roberto Livi}
\affil{Space Sciences Laboratory, University of California, Berkeley, CA 94720-7450, USA}

\author[0000-0003-3112-4201]{Robert J. MacDowall}
\affil{Solar System Exploration Division, NASA/Goddard Space Flight Center, Greenbelt, MD, 20771}

\author[0000-0001-6172-5062]{Milan Maksimovic}
\affil{LESIA, Observatoire de Paris, Universit\'{e} PSL, CNRS, Sorbonne Universit\'{e}, Universit\'{e} de Paris, 5 place Jules Janssen, 92195 Meudon,France}

\author[0000-0003-1191-1558]{David M. Malaspina}
\affil{Laboratory for Atmospheric and Space Physics, University of Colorado, Boulder, CO 80303, USA}

\author[0000-0002-2587-1342]{Juan Carlos Mart\'inez Oliveros}
\affil{Space Sciences Laboratory, University of California, Berkeley, CA 94720-7450, USA}

\author[0000-0001-6449-5274]{Nicole Meyer-Vernet}
\affil{LESIA, Observatoire de Paris, Universit\'{e} PSL, CNRS, Sorbonne Universit\'{e}, Universit\'{e} de Paris, 5 place Jules Janssen, 92195 Meudon,France}

\author[0000-0002-9621-0365]{Michel Moncuquet}
\affil{LESIA, Observatoire de Paris, Universit\'{e} PSL, CNRS, Sorbonne Universit\'{e}, Universit\'{e} de Paris, 5 place Jules Janssen, 92195 Meudon,France}

\author[0000-0002-7728-0085]{Michael Stevens}
\affil{Smithsonian Astrophysical Observatory, Cambridge, MA 02138, USA}

\author[0000-0002-7287-5098]{Phyllis Whittlesey}
\affil{Space Sciences Laboratory, University of California, Berkeley, CA 94720-7450, USA}



\begin{abstract}

We present initial results from the Radio Frequency Spectrometer (RFS), the high frequency component of the FIELDS experiment on the Parker Solar Probe (PSP).   During the first PSP solar encounter (2018 November), only a few small radio bursts were observed.  During the second encounter (2019 April), copious Type III radio bursts occurred, including intervals of radio storms where bursts occurred continuously. % Inner heliospheric observations from PSP offer the chance to observe radio emission close to the radio source, and observe signals which are less affected by scattering and refraction than those observed at 1 AU.  
In this paper, we present initial observations of the characteristics of Type III radio bursts in the inner heliosphere, calculating occurrence rates, amplitude distributions, and spectral properties of the observed bursts.  We also report observations of several bursts during the second encounter which display circular polarization in the right hand polarized sense, with a degree of polarization of $0.15-0.38$ in the range from 8-12 MHz.  The degree of polarization can be explained either by depolarization of initially 100\% polarized $o$-mode emission, or by direct generation of emission in the $o$ and $x$-mode simultaneously.  Direct \emph{in situ} observations in future PSP encounters could provide data which can distinguish these mechanisms.

\end{abstract}

\keywords{Solar radio emission (1522), Radio bursts (1339), Heliosphere (711), Space vehicle instruments (1548)}


\section{Introduction}\label{sec:intro}

Type III radio bursts are signatures of electrons accelerated in solar flares and propagating throughout the heliosphere \citep{2014RAA....14..773R}.  Type IIIs were observed and classified in the early days of solar radio observations \citep{1950AuSRA...3..387W}, distinguished from other types of radio emission by their characteristic frequency drift rate.
The basic mechanism of Type III emission was proposed by \citet{1958SvA.....2..653G}: an accelerated electron beam becomes dispersed in velocity as the electrons propagate outwards from the sun, generating bump-on-tail distribution functions which are unstable to the growth of Langmuir waves at the local plasma frequency $f_{pe}$.  The Langmuir waves are then mode converted to electromagnetic radiation at $f_{pe}$ and $2f_{pe}$, which can be remotely observed by ground and space based observatories \citep[][and references therein]{2017RvMPP...1....5M}.  Spacecraft observations of Type III radio bursts have directly measured the Langmuir waves and the electron beams associated with the radio emission \citep{1976Sci...194.1159G,1980ApJ...236..696K,1981ApJ...251..364L,1998ApJ...503..435E}.

Type III radio bursts have been observed to be partially circularly polarized using ground-based observations \citep{1980A&A....88..203D,1985srph.book..289S,2013ApJ...775...38S}.  \citet{1980A&A....88..203D} studied Type IIIs from 24-220 MHz and found an average circular polarization fraction of 0.35 for the fundamental component and 0.11 for the harmonic component.  At the lower frequencies observed by space-based radio instruments, observations of circular polarization are uncommon \citep{2019arXiv190103599C}, although there are a few observations of polarized Type III radio bursts \citep{1980A&A....91..311H}  and Type III storms \citep{2007SoPh..241..351R}.

This paper presents initial radio observations from the Parker Solar Probe (PSP) spacecraft.  The spacecraft and instrument are described in Section \ref{sec:data}, and an overview of observations is presented in Section \ref{sec:rfs_perform}.  The statistics of Type IIIs observed by PSP are presented in Section \ref{sec:stats}, and measurements of circular polarization are shown in Section \ref{sec:polarization}.  The results are \replaced{discussed}{summarized} and prospects for future observations later in the PSP mission are discussed in Section \ref{sec:discussion}.

\section{Data}\label{sec:data}

The PSP spacecraft \citep{2016SSRv..204....7F} was launched in August 2018, with a mission to study the physics of the inner heliosphere and solar corona using \emph{in situ} and remote sensing observations.  The FIELDS experiment for PSP \citep{2016SSRv..204...49B} provides the electric and magnetic measurements for the mission.  

The concept of operations for PSP divides each orbit into an encounter and a cruise phase.  During encounter phase, when the spacecraft is within 0.25 AU (54  $R_\odot$) from the sun, all instruments are on continuously and record data at high rates.  During cruise phase, instruments are on intermittently (due to power constraints and spacecraft activities) and record data at reduced rates.

Data presented in this paper come from the first two PSP solar encounters (E01 and E02).  Perihelion for E01 occurred on 2018 Nov 6, and for E02 on 2019 April 4.  For both E01 and E02, the perihelion distance was approximately 35.7 $R_\odot$, and the encounter phase lasted for approximately 5.7 days before and after perihelion.  %All PSP instruments operated nominally during E01 and E02.

The FIELDS magnetic field sensors consist of two fluxgate magnetometers (MAG) and a search coil magnetometer (SCM), with all three sensors mounted on a boom extending behind the spacecraft.  The FIELDS electric field sensors consist of four monopole electric field antennas (V1-V4), each 2 m long, mounted near the edge of the PSP heat shield, and a fifth (V5) dipole, 21 cm long, mounted on the magnetometer boom. 

These sensors are used as inputs to receivers within the FIELDS Main Electronics Package (MEP).  The FIELDS radio observations are made by the Radio Frequency Spectrometer (RFS) \citep{2017JGRA..122.2836P}, a dual channel receiver and spectrometer with a bandwith of 10.5 kHz-19.2 MHz.  RFS reduced data products are produced in two sub-bandwidths, the Low Frequency Receiver (LFR) and the High Frequency Receiver (HFR), with the LFR frequency range from 10.5 kHz-1.7 MHz, and the HFR \added{range} from 1.3 MHz-19.2 MHz.  The RFS can measure signals from the V1-V4 antenna preamplifiers and one axis of the SCM.  The electric field inputs for RFS can be configured to measure monopole or dipole signals, with a dipole measurement consisting of the difference between any two antennas.

\section{RFS Initial Performance}\label{sec:rfs_perform}

During E01 and E02, the RFS input channels (for both HFR and LFR spectra) were set to the two pairs of crossed dipoles, V1-V2 and V3-V4.  Auto-correlation and cross-correlation spectra were recorded continuously during both encounters at a cadence of 1 spectrum per $\sim$7 seconds.  During encounter periods, the FIELDS instrument is on continuously.  Outside of encounter, the FIELDS instrument is on whenever possible, taking data at reduced ``cruise mode'' rates.  During cruise mode, the RFS cadence for HFR and LFR auto and cross spectra is 1 spectrum per $\sim$56 seconds.

The two encounters were remarkably different in terms of solar radio activity.  During E01, only a few small radio bursts were observed during the entire encounter, while during E02 an active region on the sun produced copious small-to-medium sized flares, producing RFS signatures in the form of Type III radio bursts.  Comparisons of SDO/AIA and STEREO/EUVI images are consistent with the radio emissions observed in Encounter 2 originating primarily from active region NOAA 12738, at a Carrington longitude of $\sim$300$^{\circ}$.

\begin{figure*}
    %\centering
    \includegraphics[width=\textwidth]{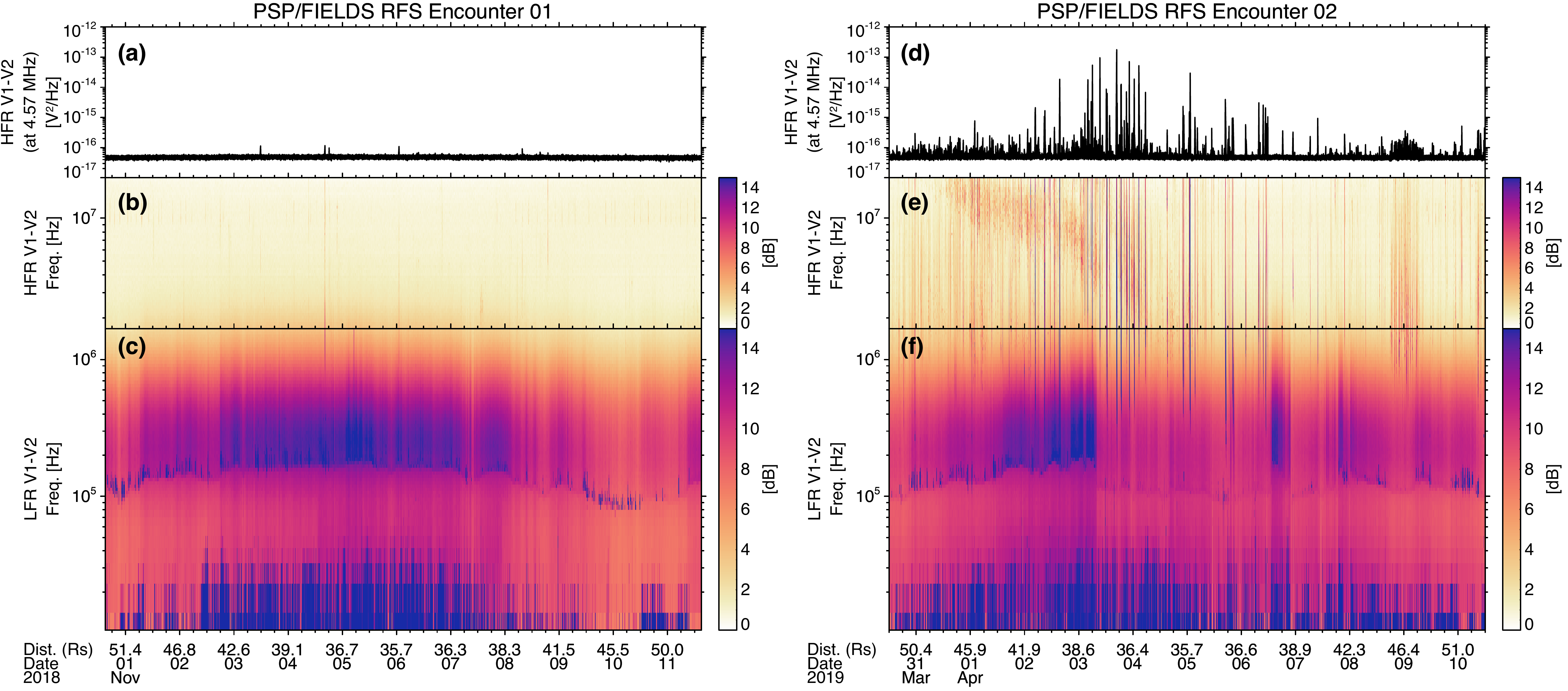}
    \caption{Comparison of RFS observations from PSP Encounters 1 and 2.  The two encounters were radically different in terms of radio emission, with Encounter 1 extremely quiet aside from a few small events, and Encounter 2 filled with numerous Type III radio bursts.}
    \label{fig:encounters}
\end{figure*}

Figure \ref{fig:encounters} shows a broad overview of the first two PSP encounters as observed in the RFS data.  For each encounter, data is shown for an 11 day interval centered on perihelion.  The left panels (Fig. \ref{fig:encounters}a-c) show E01 data.  The absence of any significant radio activity is evident in the top two panels, which shows a cut through the HFR spectrogram at 4.57 MHz (Fig. \ref{fig:encounters}a) and the full HFR spectrogram (Fig. \ref{fig:encounters}b).  The several small peaks which are visible in Fig. \ref{fig:encounters}a correspond to a few weak Type III radio bursts and intervals of Jovian emission which occurred during E01.

The bottom panel shows the full LFR spectrogram (Fig. \ref{fig:encounters}c).  At frequencies up to several hundred kHz, the LFR spectrogram is dominated by the \emph{in situ} signals from shot noise and quasi-thermal noise (QTN) \citep{1989JGR....94.2405M}.  The QTN feature at $\sim$90-150 kHz is the plasma peak, which allows RFS to make an accurate absolute determination of $f_{pe}$ and therefore the total electron density.  The prominence of the plasma peak depends on the ratio of Debye length to antenna length, and for the 2 m FIELDS antennas the peak is well resolved when $f_{pe} \gtrsim 90$ kHz.  For a detailed analysis of the QTN spectrum during the first two PSP encounters, see \citet{psp_crossref_moncuquet}.

Also visible in the LFR spectrogram are low frequency waves (observed from 10-30 kHz in LFR)  which are correlated with near-$f_{ce}$ waves observed at lower frequencies  \citep{psp_crossref_malaspina}, and large amplitude electrostatic Langmuir waves near the plasma frequency, which are evidence for electron beams in the inner heliosphere \citep{psp_crossref_bale}.  

The right panels of Figure \ref{fig:encounters} show the same data products for E02.  During E02, the sun was much more active than during E01, as is apparent in the HFR time series (\ref{fig:encounters}d) and spectrogram (\ref{fig:encounters}e).  %The increase in activity was likely associated with a northern hemisphere active region (NOAA 12738), the most prominent active region present on the Sun close to PSP perihelion 2.
In radio, the E02 solar activity is dominated by Type III radio bursts, which are the strongest emissions in the HFR frequency range.  %Typical Type IIIs observed during PSP E02 have a drift duration of $\sim$5 minutes in the HFR frequency range, and are therefore well resolved by the $\sim$7 second cadence of the HFR and LFR observations.  
Multiple Type IIIs occurred on a daily basis throughout E02, reaching a peak in intensity during $\sim$April 3-4.  %Later in the encounter, a less intense (in amplitude) but more continuous series of Type III bursts was present from April 8 18:00-April 9 8:00.
As in E01, the LFR spectrum in E02 (\ref{fig:encounters}f) shows the plasma line and electrostatic waves.  The drop in density on April 3 is consistent with measurements from the SWEAP electron and ion detectors  \citep{psp_crossref_halekas}.

After E02, a separate, nearly-continuous storm of radio bursts lasting many days occurred in mid to late April, with the most intense period on April 16-19.  During this storm interval, the PSP spacecraft had daily contacts with the NASA Deep Space Network (DSN), for transmission of encounter data the ground.  While the spacecraft is transmitting data, the instruments are turned off, so the PSP observations of this burst storm contain large daily data gaps.  This interval is discussed further in Sections \ref{sec:stats} and \ref{sec:polarization}.

\begin{figure*}
    %\centering
    \includegraphics[width=\textwidth]{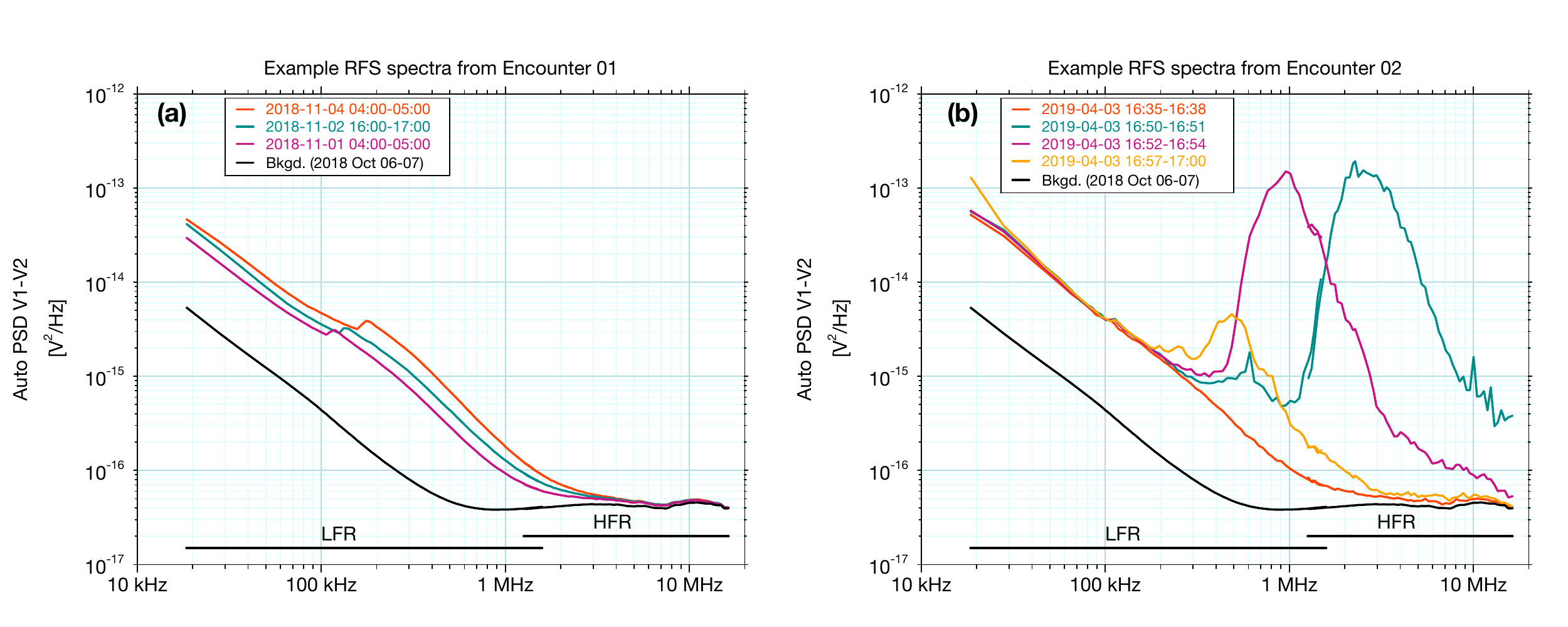}
    \caption{Example RFS spectra from the first two PSP encounters.  The left plot, from Encounter 1, shows typical quiet time RFS auto spectra, displaying the shot noise spectrum and the plasma peak in the LFR frequency range, and the galactic radio background in the HFR range.  The Encounter 2 plot on the right shows typical spectra observed during a Type III radio burst.}
    \label{fig:rfs_spectra}
\end{figure*}

The spectrogram panels in Figure \ref{fig:encounters} are presented in units of dB above background.  This presentation was chosen (rather than presenting the data in the V$^2$/Hz units of spectral density) to facilitate the display of large Type III signals without erasing faint features like the plasma line and weak radio bursts, and to avoid the spectrogram color scale being dominated by the shot noise spectrum in the LFR.  The background measurement used to produce the plot is based on spectra observed during a quiet period on 2018 October 6-7.  This background spectrum, as well as several example spectra from intervals during E01 and E02, is shown in Figure \ref{fig:rfs_spectra}.

Figure \ref{fig:rfs_spectra}a shows spectra from several quiet intervals observed during E01.  The falling, approximately power law spectrum which predominates in the LFR frequency range is a combination of shot noise and QTN.  Each E01 interval shown in Fig. \ref{fig:rfs_spectra}a is an average of all RFS Channel 0 (V1-V2) auto spectra observed over 1 hour of undisturbed solar wind, with no large fluctuations in density and no noticeable radio emission.  The small rise in each spectrum at $\gtrsim100$ kHz is the plasma peak.  Both the level of the shot noise/QTN spectrum and the frequency of the plasma peak increase with increasing density as PSP approaches close to the Sun.

In the HFR frequency range, the shot noise/QTN signal decreases below the signal from the galactic synchrotron spectrum \citep{1978ApJ...221..114N}, which is the smallest signal measured by the RFS and is on the same order as the RFS preamp/receiver input noise.  The level of the galactic synchrotron spectrum enables an accurate measurement of the effective antenna length for spacecraft radio receivers \citep{2009RaSc...44.4012E,2011RaSc...46.2008Z}.  Preliminary calibrations for the RFS indicate an effective length of $\sim$3 m when the crossed dipole antennas (V1-V2 and V3-V4) are used as inputs to the two RFS channels.

Figure \ref{fig:rfs_spectra}b shows several shorter intervals (1-3 minutes) during the more active E02, on April 3.  The spectra are less smooth than those in \ref{fig:rfs_spectra}a due to the shorter averaging intervals.  One interval, from 16:35-16:38, shows a similar undisturbed profile as those shown in \ref{fig:rfs_spectra}a.  The other spectra are taken during a typical large radio burst observed during Encounter 2.  As seen at 1 AU, peak radio intensity occurs at frequencies of $\sim 1$ MHz \citep{2014SoPh..289.3121K}.  Detailed comparison of radio burst intensity profiles in the inner heliosphere  to those observed at 1 AU is beyond the scope of this work, but is likely to be a fruitful area of research as PSP approaches closer to the Sun--especially when the launch of Solar Orbiter adds another point to the available observations.

In Figure \ref{fig:rfs_spectra}, the frequency coverage of the LFR and HFR data products is shown below the spectra.  The lowest frequency bin in LFR is not shown, since it is dominated by the plasma waves described by \citet{psp_crossref_malaspina} which vary over the averaging intervals used in the figure.  Several of the highest HFR frequency bins above 16 MHz are also not included, since they contain some aliased power from above the Nyquist frequency of $19.2$ MHz.

\begin{figure*}
    %\centering
    \includegraphics[width=\textwidth]{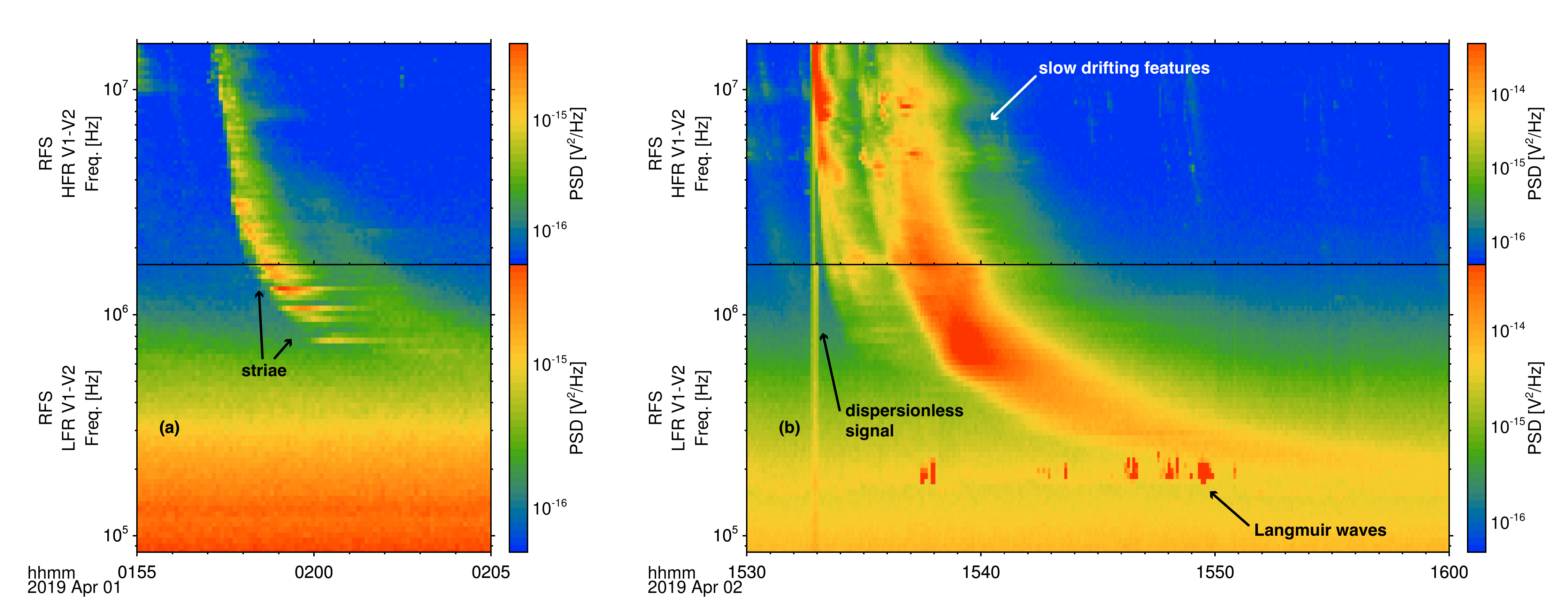}
    \caption{Examples of radio bursts observed during PSP Encounter 2. The left plot (a) shows a Type III with striations due to density fluctuations, and the right plot (b) \replaced{shows a strong and complex event featuring signatures of flare photons, multiple interacting Type IIIs, and slowly drifting Type II-like features.}{shows a strong event with a vertical (dispersionless) signal near the start and slowly-drifting Type II-like features.}}
    \label{fig:striated_and_large_burst}
\end{figure*}

Figure \ref{fig:striated_and_large_burst} shows examples of events observed by RFS during Encounter 2.  The burst observed on April 1, shown in Fig. \ref{fig:striated_and_large_burst}a, is a Type IIIb burst, featuring frequency structures known as striae \citep{1972A&A....20...55D}.  The striae are associated with density inhomogeneities in the source region of the burst \citep{2017NatCo...8.1515K,2018SoPh..293..115S}.  Fig. \ref{fig:striated_and_large_burst}b shows a large amplitude, \deleted{complex} event observed on April 2, \added{with slowly drifting Type II-like features from $\sim$5--10 MHz.}  \replaced{The vertical feature near the start of the event is likely a direct signal of flare photons impacting the antennas, generating a short-lived change in the antenna potential.}{The vertical feature near the start of the event is dispersionless, to the time resolution of the measurement, and extends
below the local plasma frequency, indicating that it cannot be freely propagating radio emission.  We suggest that this dispersionless signal corresponds to a change in the relative floating voltage of the antennas and spacecraft (and therefore system gain) due to photoemission from the impulsive UV emission associated with the source flare.}  \deleted{The radio observations from this complex event show interactions of multiple separate radio bursts, features with drift rates considerably slower than Type III rates, and eventual coalescence of separate features into a single burst below $\sim$1 MHz.  Complex events of this type may indicate a similarly complex magnetic structure, with only some open field lines for electron beams to escape into interplanetary space \citep{1999GeoRL..26..397R,2000ApJ...530.1049R}.}   Electrostatic Langmuir waves are observed near the plasma frequency and could possibly be associated with the radio burst--however, we note that PSP frequently observes Langmuir waves in the inner heliosphere in the absence of radio emission \citep{psp_crossref_bale}.

Before proceeding to detailed analysis of the radio bursts observed in E02, we note some limitations to these initial results.  The calibration of the antenna effective vectors has not yet been completed, so we present results in units of power spectral density (PSD), $\mathrm{V^2/Hz}$, rather than physical units of sfu or $\mathrm{W/m^2/Hz}$.  However, on-orbit measurements during spacecraft maneuvers have indicated that the RFS receiver is sufficiently sensitive for observations of the galactic synchrotron spectrum, making an accurate absolute calibration in the manner of \citet{2009RaSc...44.4012E} and \citet{2011RaSc...46.2008Z} possible.  

We also note that in the statistical analysis presented in the following section, the measured PSD for observed bursts has been  radially scaled by a factor of $(R/R_0)^2$, where $R_0 = 35.7 R_{\sun}$ is the perihelion radius and $R$ is the radial distance of PSP from the sun.  This rough scaling gives a first order correction to allow for comparison of burst amplitudes over the encounter.  For the radial distance range shown in Figure \ref{fig:encounters}, this is a factor of $\sim$2 between the closest and most distant samples in the spectrogram ($(54\:R_{\sun}/37.5\:R_{\sun})^2$).  A more accurate scale factor would correct for the radial distance of the emission, which over the frequency range of 1--16 MHz corresponds to radial distances of 1.6--5 $R_{\sun}$ \citep{1998SoPh..183..165L}.  This second order correction could increase or decrease the observed amplitude relative to the source amplitude, depending on the location of the radio emission.  As the PSP perihelion distance decreases throughout the mission, the importance of this correction will increase, and will likely require detailed goniopolarimetric/direction finding analysis for determination of the source location \citep{2008SSRv..136..549C,2014SoPh..289.4633K}.

Due to these limitations of the current data set, in this work we restrict analysis to quantities (power law indices, waiting time distribution, and relative polarization) that do not depend on absolute determination of amplitude.

\section{Statistics of Type III Radio Bursts During E02}\label{sec:stats}

\begin{figure*}%[htbp]
    %\centering
    \includegraphics[width=\textwidth]{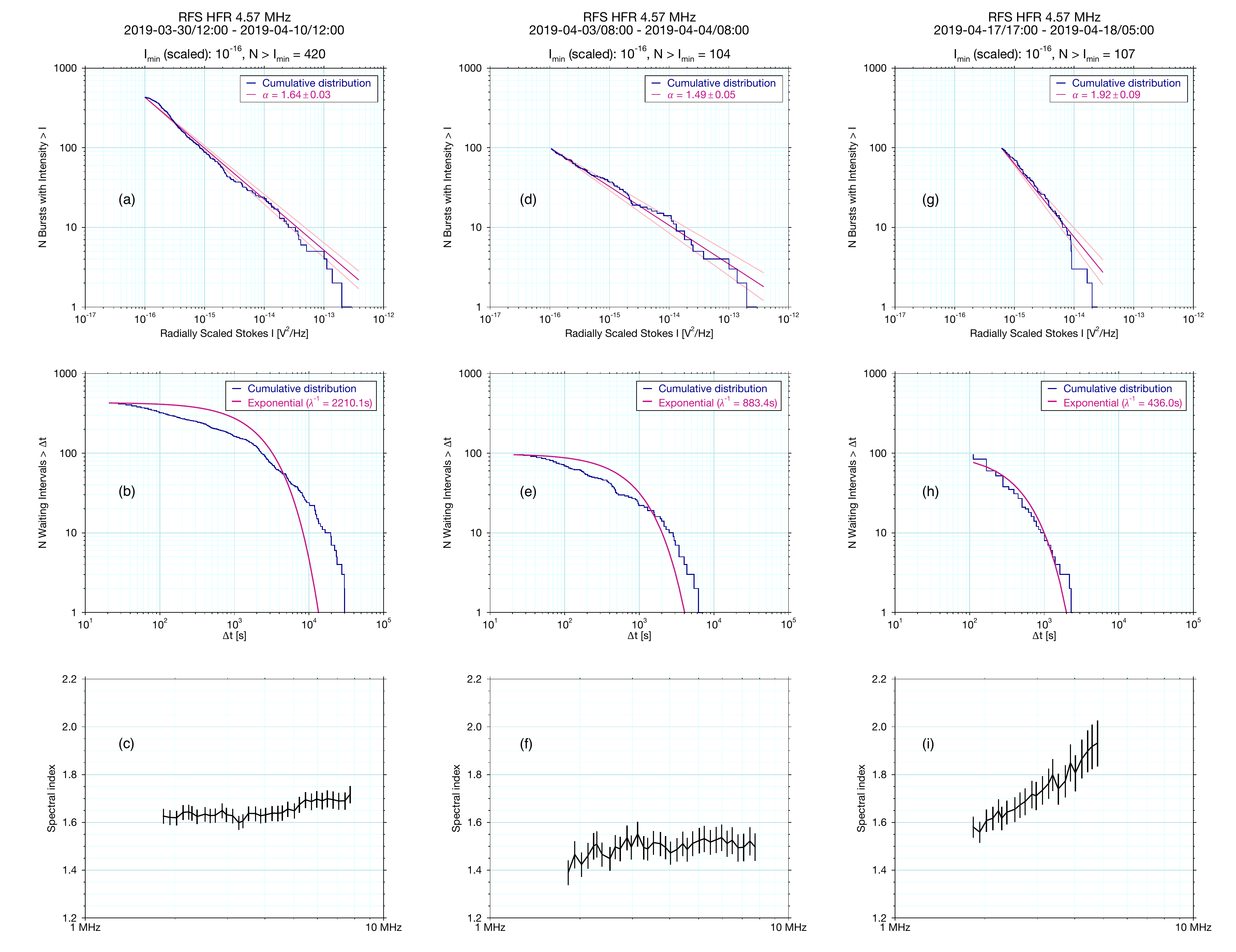}
    \caption{Statistics of Type III radio burst occurrence during and after PSP Encounter 2.  The three columns show statistics for an interval covering all of Encounter 2 (a-c), the 24 hours showing the most intense burst activity (d-f), and a post-encounter interval during a Type III storm (g-i).  Each column shows the distribution of burst intensity (top), the waiting time distribution (center), and the variation in power law index $\alpha$ with frequency (bottom).}
    \label{fig:statistics}
\end{figure*}

Figure \ref{fig:statistics} shows some basic statistical parameters of Type III radio bursts observed during E02.  Three time intervals are presented: in the first column (Figure \ref{fig:statistics} a-c), statistics are shown for the entire encounter, using the same time range as in Figure \ref{fig:encounters}.  The second column covers a single 24 hour period from 2019 April 03/08:00 to 2019 April 04/08:00, which was the most intense period of radio burst activity during E02.  The final column covers an interval during the mid-April Type III storm, when the RFS instrument was turned on and making observations at a reduced cruise mode rate.  Figure \ref{fig:burst_storm} shows HFR observations during this portion of the storm.

Individual bursts were detected in the RFS spectrograms by identifying maxima in cuts through the spectrogram at a given frequency.  Bursts were required to be above a threshold value to eliminate random fluctuations, the effects of high frequency electron thermal noise \citep{psp_crossref_maksimovic}, and observed Jovian emission.  For all analyzed intervals, a minimum scaled PSD of $I_{\mathrm{min}} = 10^{-16}\: \mathrm{V^2/Hz}$ was used as a threshold.  To calculate intensity $I$, we used the sum of the $V1-V2$ and $V3-V4$ cross dipoles.   

A lower threshold for $I$ would result in a count of $\gtrsim$ 1000 Type III radio bursts observed during E02, but would be contaminated by some non-Type III sources described above.  Above the threshold, there are 420 bursts during the 11 days of E02, 104 bursts during the most intense 24 hour period on April 3/08:00 to April 4/08:00, and 107 during the 12 hour segment of the Type III storm from April 17/17:00 to April 18/05:00.

The top panels of Figure \ref{fig:statistics} show the distribution of PSD for all bursts $> I_{\mathrm{min}}$ at a sample HFR frequency of 4.57 MHz.  Following the technique used in \citep{2010ApJ...708L..95E}, we plot the cumulative distribution of burst intensity $I$ and calculate a power law index $\alpha$ using the maximum likelihood method.  From an intensity distribution $f(I) \propto I^{-\alpha}$, $\alpha$ and $\sigma_\alpha$ can be computed as \citep{2004ApJ...609.1134W}:

\begin{equation}
    \alpha = \frac{N}{\Sigma_{i=1}^N \mathrm{ln}(I_i/I_{\min})} + 1
\end{equation}

\begin{equation}
    \sigma_\alpha \approx (\alpha - 1) N^{1/2}
\end{equation}

In the cumulative distribution plots shown in Fig. \ref{fig:statistics}(a,d,g), a power law index of $\alpha$ corresponds to a slope of $-\alpha + 1$.  For each interval, the intensity distribution follows a power law reasonably well within the expected uncertainty.  The power law index is flattest ($\alpha = 1.49\pm0.05$) during the time period of the most intense bursts (Figure \ref{fig:statistics}d), somewhat steeper ($\alpha = 1.64\pm0.03$) over the entire encounter (Figure \ref{fig:statistics}a), and steepest ($\alpha = 1.92\pm0.03$) during the burst storm 
(Figure \ref{fig:statistics}g).  The power law index during the storm agrees reasonably well with the index of $2.10\pm0.05$ observed by \citet{2010ApJ...708L..95E} at 5.025 MHz, while the steeper index observed during the non-storm encounter period agrees with the index of $1.69$ found by \citet{1976SoPh...46..465F}.

Cumulative waiting time distributions are shown in Figure \ref{fig:statistics}(b,e,h).  For reference, a model waiting time distribution curve corresponding to a simple Poisson process is plotted along with the cumulative distributions.  This model is given by:

\begin{equation}
    N(t_{\mathrm{wait}} > \Delta t) \propto \exp^{-\lambda \Delta t}
\end{equation}

where $\lambda$ is the rate of occurrence of bursts during the interval.  The Type III storm interval appears to be consistent with a single continuous Poisson process, while the waiting time distributions during encounter display more structure, corresponding to distinct intervals of different activity.  %(This is also consistent with the variation in $\delta$ between the entire E02 interval and the subset on April 3-4.)  
A Bayesian block analysis of the type performed by \citet{2010ApJ...708L..95E} could be used to separate sub-intervals during the encounter, with each distinct sub-interval associated with an independent rate of burst activity.

The variation of spectral index $\alpha$ with frequency is shown in Figure \ref{fig:statistics}(c,f,i).  We have limited the frequency range in these plots to $f<8$ MHz, because there is an intermittent source of noise (possibly instrument or spacecraft-generated) which occurs sporadically throughout the encounter for a single spectra measurement at frequencies above 8 MHz.  In the case of the Type III storm, we have further limited our analysis to below 5 MHz, because above 5 MHz the radio bursts have a duration lower than the cadence of the observations.  We also limit the frequency range to $f>1.8$ MHz, to avoid the effects of the high frequency tail of the QTN spectrum during encounter.

During the encounter period the spectral index remained at $\alpha \sim 1.6-1.7$ from $1.8-8$ MHz, with a possible steepening at higher frequencies.  During the 24 hour peak intensity period, the spectral index was slightly lower at $\alpha \sim 1.4-1.5$, and nearly constant over the entire frequency range.  In contrast, and also in contrast to the storm analyzed by \citet{2010ApJ...708L..95E}, the late April storm becomes steeper with increasing frequency, from $\alpha \sim 1.6$ at 1.8 MHz to $\alpha \sim 1.9$ at 8 MHz.

Variation in $\alpha$ with frequency may be due to the location of the active region relative to the spacecraft,  the refraction of emitted radiation \citep{2012ApJ...745..187T}, and scattering of the emission as it propagates.  Multipoint measurements of directivity patterns \citep{1989A&A...217..237L,2008A&A...489..419B,2009SoPh..259..255R}, and the study of many burst intervals will be useful in determining why some periods exhibit no variation of $\alpha$ with frequency (\citet{2010ApJ...708L..95E}, the encounter periods analyzed here) and some periods exhibit definite trends (\citet{1976SoPh...46..465F}, \replaced{our cruise mode storm period}{the late April storm analyzed in this paper}).

\section{Polarization of Type III Radio Bursts}\label{sec:polarization}

In this section, we describe the circular polarization observed for a subset of radio bursts during E02.  The analysis in the previous section employed the Stokes intensity ($I$) parameter.  From the RFS auto and cross spectra products, it is also possible to calculate the Stokes $Q$, $U$, and $V$ parameters, representing the linear ($Q$, $U$) and circular ($V$) polarization:

\newcommand*{\autoa}{{V}_{12} {V}_{12}^*}
\newcommand*{\autob}{{V}_{34} {V}_{34}^*}
\newcommand*{\xspec}{{V}_{12} {V}_{34}^*}

\begin{equation}
I = \autoa + \autob
\end{equation}
\begin{equation}
Q = \autoa - \autob
\end{equation}
\begin{equation}
U = \xspec + (\xspec)^*
\end{equation}
\begin{equation}
iV = \xspec - (\xspec)^*
\label{eqn:circular}
\end{equation}

where $\autoa$ and $\autob$ are the RFS auto spectra \added{from the $V1-V2$ and $V3-V4$ channels}, and $\xspec$ is the RFS cross spectrum.    Note that we use the convention where right-hand circular (RHC) polarization indicates an electric field vector rotating clockwise when viewed from the radio source in the direction of propagation towards the spacecraft.  In Equation \ref{eqn:circular}, $V < 0$ corresponds to RHC polarization.  Although we can in principle calculate all Stokes polarization parameters, we only present $V$ here because the linear polarization Stokes parameters are affected by Faraday rotation \citep{1985srph.book..289S} so any source linear polarization ($U$, $Q$) is unlikely to be measurable remotely. 

In general, calculation of the Stokes parameters requires correction by the Mueller matrix, which compensates for instrumental effects on observed polarization of radio sources \citep[e.g.][]{2001PASP..113.1274H}.  For the case of short dipole antennas such as the 2 m FIELDS antennas, the effective length vectors of the antennas are real-valued and this correction is substantially simplified \citep{2011pre7.conf...13L}.  Under the assumption that the true linear polarization of the Type III emission is negligible, and noting that $Q \approx U \approx 0$ during the burst intervals where circular polarization $V$ was observed, Equation 6 of \citet{2011pre7.conf...13L} shows that a simple computation of relative circular polarization ($V/I$) is approximately valid even without detailed consideration of arrival direction or corrections for antenna non-orthogonality.  

\subsection{Polarization of individual radio bursts observed near perihelion}

\begin{figure*}
    %\centering
    \includegraphics[width=\textwidth]{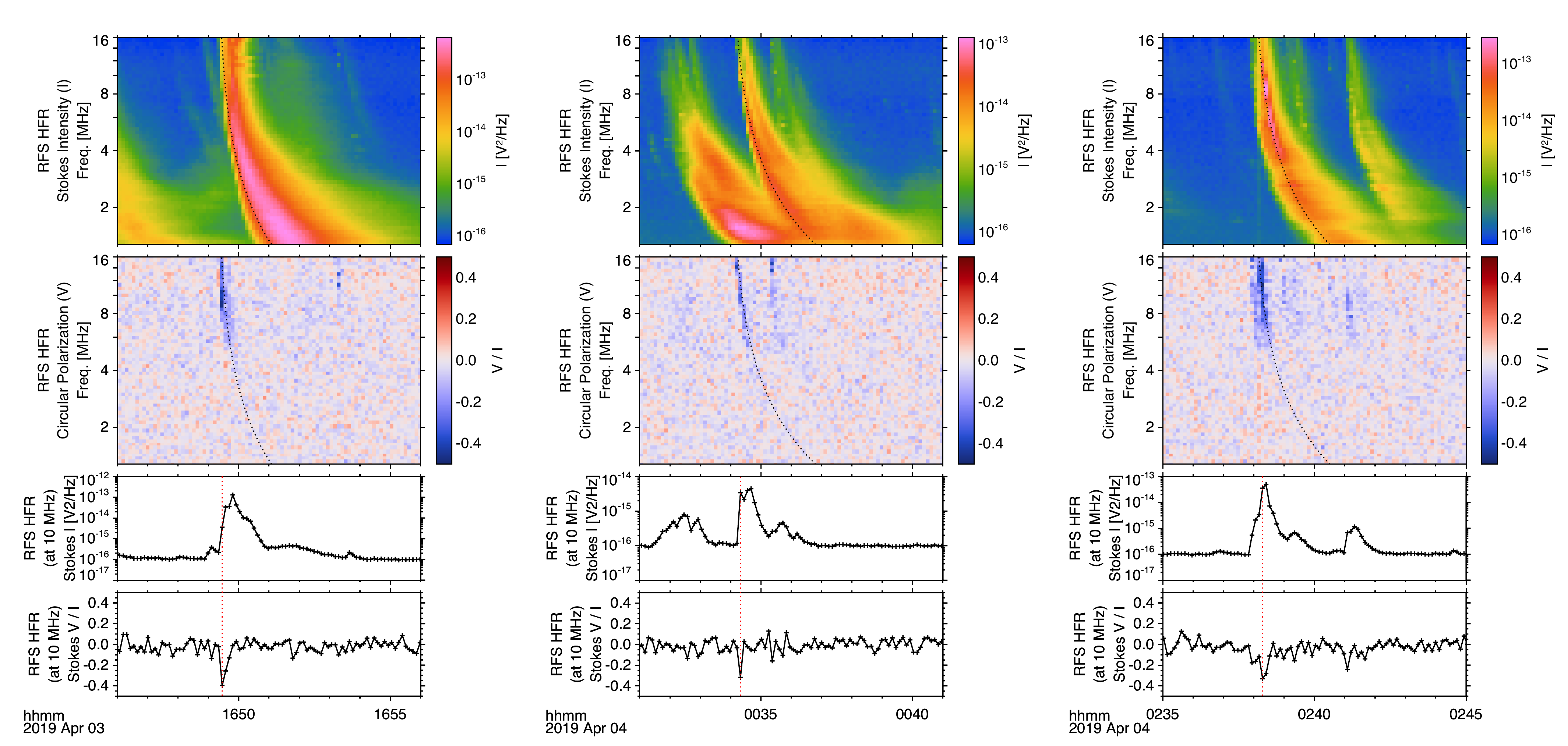}
    \caption{Example radio bursts displaying circular polarization above $\sim$6 MHz.  In each plot, the top panel shows the Stokes intensity ($I$), while the \replaced{bottom}{second} panel shows the relative circular polarization Stokes $V/I$.  Negative $V$, in blue, indicates right hand circular polarization.  In each spectrogram plot, the dotted line indicates the time profile of the leading edge of the burst.  \deleted{The burst starting on April 4 at $\sim$00:35 shows apparent fundamental-harmonic ($F-H$) structure, with circular polarization apparent in the $F$ component.} \added{The bottom two panels show time profiles of $I$ and $V/I$ at a frequency of 10 MHz, showing how the polarization is localized near the leading edge of the burst and is absent at later times.  The time of maximum circular polarization is indicated in these panels with a red dotted line.}}
    \label{fig:example_burst}
\end{figure*}

Figure \ref{fig:example_burst} shows \added{$I$ and $V/I$ spectrograms and time profiles (at 10 MHz) for} three radio bursts exhibiting circular polarization \deleted{measured by RFS in E02} (out of seven bursts identified \added{in E02} with clear signatures of circular polarization).  In each burst, RHC polarization is present for a short time near the leading edge of the burst, at frequencies above $\sim$6 MHz.    Table \ref{tab:bursts_table} presents the time of the leading edge of the burst near the top and bottom of the HFR range ($t_0$ and $t_1$), and the fraction of circular polarization $V/I$, observed for each of the seven individual bursts analyzed in this study.  In addition to the seven bursts analyzed here, numerous other bursts presented weaker, marginally distinguishable signals.  For example, in Figure \ref{fig:example_burst}, the weaker Type IIIs starting at approximately 00:35:30 and 02:41:00 display faint possible signals of circular polarization.

\begin{table}[htbp]    \centering
    \begin{tabular}{c|c|c}
    $t_0$ (16 MHz) & $t_1$ (1.3 MHz) & $V/I$ (8-12 MHz) \\
April 03 09:22:38 & April 03 09:25:18 & -0.28 \\
April 03 16:49:26 & April 03 16:51:05 & -0.22 \\
April 03 18:49:04 & April 03 18:51:25 & -0.24 \\
April 03 22:22:59 & April 03 22:25:14 & -0.15 \\
April 04 00:34:14 & April 04 00:36:48 & -0.19 \\
April 04 02:38:10 & April 04 02:40:30 & -0.38 \\
April 04 05:35:04 & April 04 05:37:23 & -0.16
    \end{tabular}
    \caption{Time and circular polarization for seven radio bursts observed during Encounter 2.  The times $t_0$ and $t_1$ correspond to the times of the leading edge of the radio burst.  The dotted frequency profiles shown in Figure \ref{fig:example_burst} are derived from these times.}
    \label{tab:bursts_table}
\end{table}

\subsection{Polarization of Type III storm}\label{sec:storm}

\begin{figure}
    %\centering
    %\includegraphics[width=\textwidth]{polarized_storm_v2.pdf}
    \plotone{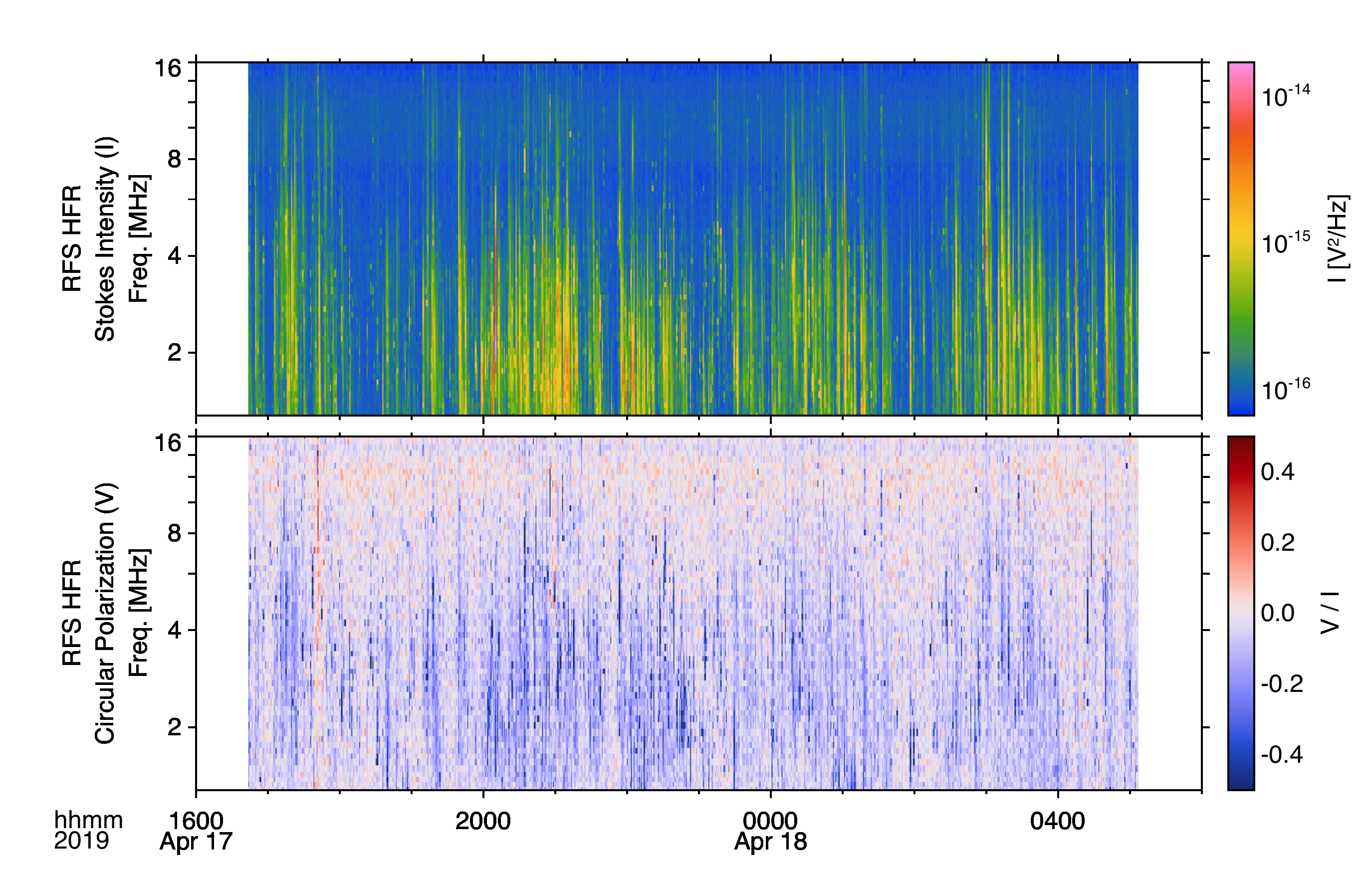}
    \caption{Type III storm with circular polarization.  Like the individual events in the previous section, the sense of the polarization is RHC.}
    \label{fig:burst_storm}
\end{figure}

As mentioned in previous sections, a days-long Type III storm was observed by RFS during mid-late April.  At this time, the FIELDS instruments were in cruise mode, recording spectra approximately once per 56 seconds.  An example interval is shown in Figure \ref{fig:burst_storm}.  Unlike any of the active periods observed during the encounter, this storm exhibited significant circular polarization throughout the storm period, associated with nearly every observed burst.  As in the bursts observed during encounter, the polarized emission is overwhelmingly in the RHC sense.  The average degree of polarization increases slightly with frequency, from about $-0.1$ at 1 MHz to $-0.2$ at 5 MHz (above 5 MHz, the cruise mode cadence is too slow to clearly distinguish individual bursts).  

\subsection{Analysis of Polarization Observations}

\begin{figure*}
    %\centering
    \includegraphics[width=\textwidth]{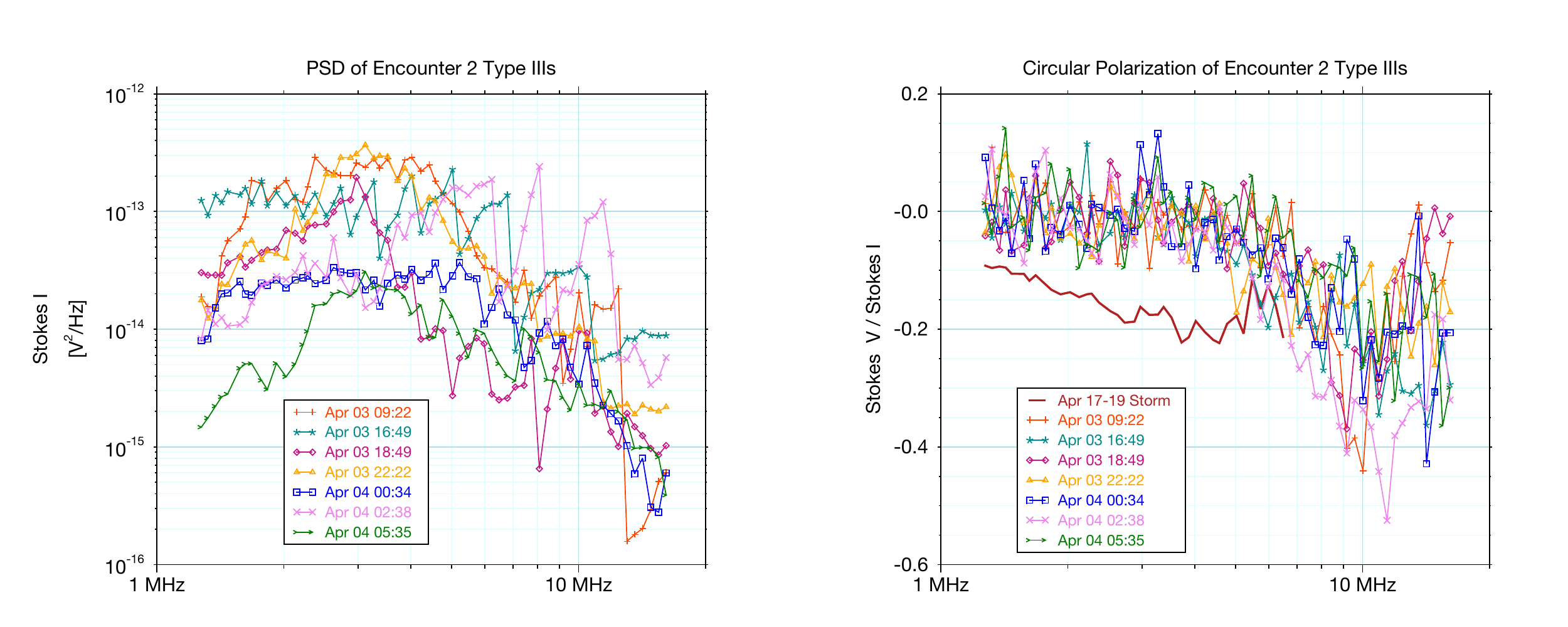}
    \caption{Stokes $I$ and $V/I$ profiles of Type III radio bursts observed during PSP Encounter 2.  For each individual burst, the profiles are derived from a frequency-dependent cut through the spectrogram, as shown in Figure \ref{fig:example_burst}.  The average $V/I$ for the storm period shown in Figure \ref{fig:burst_storm} is shown along with the individual cuts.}
    \label{fig:burst_cuts}
\end{figure*}

Figure \ref{fig:burst_cuts} shows profiles of Stokes intensity $I$ and circular polarization $V/I$ for the bursts in Table \ref{tab:bursts_table}.  The profiles are based on cuts through the spectrogram plots at the times defined in Table \ref{tab:bursts_table}, which are shown as dotted lines in Figure \ref{fig:example_burst}.  The intensity profiles are consistent with typical Type III profiles observed at 1 AU \citep{2014SoPh..289.3121K}.  \replaced{However, it}{It} is not immediately apparent why only a small fraction of bursts observed during E02 showed signatures of circular polarization.  From comparing Figure \ref{fig:burst_cuts} with Figure \ref{fig:statistics}d, it is clear that the seven bursts showing circular polarization are relatively high in amplitude.  However, many comparably large bursts, including the two examples in Figure \ref{fig:striated_and_large_burst}, did not show signs of circular polarization, \added{and the small number of observations $(N=7)$ does not permit a detailed quantitative analysis of the relation between burst intensity and degree of polarization}.

For each of the seven analyzed bursts, Stokes $V/I$ increases from a value $\sim$0 below 4 MHz, to a value of $-0.15$ to $-0.38$ in the range from 8-12 MHz, indicating partial RHC polarization.  This agrees reasonably well with previous observations at higher frequencies \citep{1980A&A....88..203D}.

The circularly polarized component of these bursts likely corresponds to fundamental ($F$) emission at $f_{pe}$, and not harmonic ($H$) emission at $2 f_{pe}$.  \deleted{For most of the bursts, there are not obvious distinct $F$ and $H$ components.  However, there is one burst (starting in HFR at April 04 00:34) which does show apparent $F-H$ structure, and the circular polarization appears to follow the $F$ component.  (See Figure \ref{fig:example_burst}, center panel).  In addition, previous} \added{Previous} studies of Type III bursts at starting at slightly higher frequencies ($>24$ MHz) are consistent with the values of $V/I$ in Table \ref{tab:bursts_table} assuming they correspond to the $F$ component, while the values of $V/I$ for the $H$ component are lower \citep{1980A&A....88..203D}.

For all bursts listed in Table \ref{tab:bursts_table}, the maximum value of circular polarization $V/I$ precedes the maximum intensity $I$, and polarized signal is absent from the latter stage of the burst (see Figure  \ref{fig:example_burst}).  \added{This is also consistent with previous results \citep{1998JGR...10317223D} showing that the initial phase of radio bursts is dominated by fundamental emission, which is more strongly polarized than  harmonic emission.  After the initial period (i.e. at the peak and during the exponential decay of the burst), emission may be from the fundamental or harmonic.  If this later part of the burst is dominated by harmonic emission, then the weaker polarization signal is naturally explained by the weaker polarization observed in the harmonic component of Type IIIs \citep{1984A&A...130...39D}.  If the emission in the later part of the burst is primarily fundamental, then the initially polarized signal could be affected by scattering in the inner heliosphere.}  \deleted{A possible explanation for this is scattering of the radio signal in the inner heliosphere.}  A recent study by \citet{2018ApJ...857...82K} simulated the time profile profile of Type IIIs based on a Monte Carlo model, concluding that the exponential decay of the intensity profile could be explained by small scale density fluctuations.  The same scattering process could reduce the coherence of an initially circularly polarized signal, leaving the latter part of the burst unpolarized as measured by the observer.

Previous observations have established a correlation with the sense of circular polarization (for Type I and Type III bursts) and the polarity of the leading sunspot in the active region associated with the radio emission \citep{1984A&A...130...39D,1985SoPh...96..381S,2007SoPh..241..351R}.  This is consistent with the SDO/HMI magnetograms during the month of April, which show the active region (NOAA 12738) in the Northern hemisphere, with a leading spot with negative polarity (field pointing into the Sun) and a trailing spot with positive polarity (field pointing out from the Sun).




%\begin{thebibliography}{}
%\end{thebibliography}

\section{Discussion} \label{sec:discussion}

A major reason that the polarization of Type III radiation is of interest is that it provides a remote probe of the coronal magnetic field, where the degree of circular polarization is dependent on the strength of the field.  In the case of harmonic emission, theory predicts a direct relation between the degree of polarization and the ratio of electron gyrofrequency and the plasma frequency, $V/I \sim A(f_{B}/f_{p})$ \citep{1984A&A...130...39D}.  This relation was used by \citet{2007SoPh..241..351R} to infer the profile of the coronal magnetic field, and compare to model coronal fields \citep{1978SoPh...57..279D}.

The polarization of the fundamental component is less straightforward, but also depends on the properties of the coronal field.  Magnetoionic theory predicts that fundamental emission is generated with 100\% polarization in the $o$-mode, which is observed in Type I radio bursts \citep{2017RvMPP...1....5M}.  However, for Type III radio bursts, 100\% polarization is not observed, which implies that some depolarization mechanism must be in effect to reduce the completely polarized emission to the lower levels observed.  \citet{2006ApJ...637.1113M} has proposed reflection as a mechanism, with emission created in regions of lower density (ducts) surrounded by regions of higher density.  The radio emission would reflect off the duct boundaries, and in the process become depolarized.  An alternative explanation was proposed by \citet{2007PhRvL..99a5003K}, who developed theory and simulations to show that, under conditions consistent with solar wind observations, emission can be generated partially in the $o$ mode and partially in the $x$ mode.  Direct generation of emission in different modes could account for the observed $<$100\% polarization without need for a depolarization mechanism.

\subsection{Comparison with previous spacecraft measurements}

As discussed previously, partial circular polarization has been observed from the ground, but previous spacecraft measurements have not reported polarization signatures at frequencies greater than a few MHz.  The reason why RFS observes polarization at frequencies $\gtrsim 6$ MHz is most likely not due to observing position or any particular antenna or spacecraft geometry.  Although the intensity (in sfu or $\mathrm{W/m^2/Hz}$) for a given burst is higher in the inner heliosphere, the 2 m PSP antennas are considerably shorter than previous radio instruments such as STEREO \citep{2008SSRv..136..487B}  (6 m triaxial antennas) or Wind \citep{1995SSRv...71..231B} (100 m and 15 m tip to tip dipoles in the spin plane, and 12 m in the axial direction).  Therefore, the signal to noise ratio is not significantly higher for PSP than for previous spacecraft, at least not during the initial encounters.  The cross dipole configuration of PSP is convenient for simple computation of Stokes parameters as presented in this work, but both Wind and STEREO are fully capable of making goniopolarimetric measurements of source polarization \citep{1985A&A...153..145F,2008SSRv..136..549C,2014SoPh..289.4633K}.

The most likely reason that similar signatures of polarized Type III bursts have not been reported using STEREO or Wind data is the relatively slow cadence of the radio measurements.  For STEREO/WAVES HFR, the cadence is typically $\sim$40 seconds, likely too short to measure the circular polarization, which typically only appears in the RFS data for 1-3 spectra at 7 second cadence.  For Wind, the relevant frequency range is covered by the RAD2 receiver, which typically makes measurements at $\sim$16 seconds, which could be marginal for short-lasting circular polarization observations.  However, direction finding mode for RAD2, which is required for measuring polarization, is not always enabled.

\subsection{Absence of Associated \emph{in situ} Electron Observations}
As of the time of writing, there were no observations of \emph{in situ} electron beams associated with any of the Type III radio bursts in E02, of the type that have been \added{previously} observed \added{by other spacecraft} \deleted{ for some large events with previous spacecraft} \citep{1981ApJ...251..364L,1998ApJ...503..435E}.  This is likely due to the magnetic connectivity of PSP during E02.  The active region likely responsible for the bursts (NOAA 12738) was in the northern hemisphere, while during the encounter the radial component of the magnetic field $B_r$ was predominantly negative, indicating connection to the Sun on the southward side of the heliospheric current sheet.

Additionally, the instrument configuration on PSP for the first two encounters was not ideal for detection of the several to $\sim$10 keV electron beams which generate Langmuir waves.  This energy range is below the minimum energies of the ISOIS electron detectors \citep{2016SSRv..204..187M}, and the SWEAP electron electrostatic analyzers were configured with a maximum energy of 2 keV.  In future encounters, the maximum SWEAP energy could be extended (although at the expense of cadence or energy resolution).

The PSP perihelion distance will decrease throughout the course of the mission, eventually reaching closest approach of of 9.86 $R_\sun$ in 2024.  At that time, the Sun is expected to be at or near the maximum of the solar cycle, and conditions will be ripe for detection of a \emph{in situ} Type III in the inner heliosphere (ideally with electron beams in the key energy range).  The \emph{in situ} measurements made during such an event will provide observational tests of proposed theories of radio emission.  As an example, the \emph{in situ} density measurements in the inner heliosphere could show evidence of the ducting structures proposed by \citet{2006ApJ...637.1113M}, and high cadence waveforms recorded by the FIELDS Time Domain Sampler (TDS) instrument could measure emission wave modes directly from within a radio burst source region, to test against the predictions of \citet{2007PhRvL..99a5003K}.  These inner heliospheric measurements, combined with other ground and space-based radio observations (including Solar Orbiter after 2020) will offer new fundamental insights into the nature of solar radio emission.

\acknowledgments
The FIELDS experiment on the Parker Solar Probe spacecraft was designed and developed under NASA contract NNN06AA01C. The FIELDS team acknowledges the extraordinary contributions of the Parker Solar Probe mission operations and spacecraft engineering teams at the Johns Hopkins University Applied Physics Laboratory.  The RFS science team is immensely grateful to Dennis Seitz and Dorothy Gordon for their contributions to the analog and digital design of the RFS receiver.  MP acknowledges useful discussions with Vratislav Krupar.  SDB acknowledges the support of the Leverhulme Trust Visiting Professorship program.  Data access and processing was performed using SPEDAS \citep{2019SSRv..215....9A}.  PSP/FIELDS data is publicly available at \url{http://fields.ssl.berkeley.edu/data/}.

\bibliographystyle{aasjournal}
\bibliography{pulupa_typeiii_papers,pulupa_typeiii_crossref}

\begin{thebibliography}{}
\expandafter\ifx\csname natexlab\endcsname\relax\def\natexlab#1{#1}\fi
\providecommand{\url}[1]{\href{#1}{#1}}
\providecommand{\dodoi}[1]{doi:~\href{http://doi.org/#1}{\nolinkurl{#1}}}
\providecommand{\doeprint}[1]{\href{http://ascl.net/#1}{\nolinkurl{http://ascl.net/#1}}}
\providecommand{\doarXiv}[1]{\href{https://arxiv.org/abs/#1}{\nolinkurl{https://arxiv.org/abs/#1}}}

\bibitem[{{Angelopoulos} {et~al.}(2019){Angelopoulos}, {Cruce}, {Drozdov},
  {Grimes}, {Hatzigeorgiu}, {King}, {Larson}, {Lewis}, {McTiernan}, {Roberts},
  {Russell}, {Hori}, {Kasahara}, {Kumamoto}, {Matsuoka}, {Miyashita},
  {Miyoshi}, {Shinohara}, {Teramoto}, {Faden}, {Halford}, {McCarthy}, {Millan},
  {Sample}, {Smith}, {Woodger}, {Masson}, {Narock}, {Asamura}, {Chang},
  {Chiang}, {Kazama}, {Keika}, {Matsuda}, {Segawa}, {Seki}, {Shoji}, {Tam},
  {Umemura}, {Wang}, {Wang}, {Redmon}, {Rodriguez}, {Singer}, {Vandegriff},
  {Abe}, {Nose}, {Shinbori}, {Tanaka}, {UeNo}, {Andersson}, {Dunn}, {Fowler},
  {Halekas}, {Hara}, {Harada}, {Lee}, {Lillis}, {Mitchell}, {Argall},
  {Bromund}, {Burch}, {Cohen}, {Galloy}, {Giles}, {Jaynes}, {Le Contel}, {Oka},
  {Phan}, {Walsh}, {Westlake}, {Wilder}, {Bale}, {Livi}, {Pulupa},
  {Whittlesey}, {DeWolfe}, {Harter}, {Lucas}, {Auster}, {Bonnell}, {Cully},
  {Donovan}, {Ergun}, {Frey}, {Jackel}, {Keiling}, {Korth}, {McFadden},
  {Nishimura}, {Plaschke}, {Robert}, {Turner}, {Weygand }, {Candey}, {Johnson},
  {Kovalick}, {Liu}, {McGuire}, {Breneman}, {Kersten}, \&
  {Schroeder}}]{2019SSRv..215....9A}
{Angelopoulos}, V., {Cruce}, P., {Drozdov}, A., {et~al.} 2019, \ssr, 215, 9,
  \dodoi{10.1007/s11214-018-0576-4}

\bibitem[{{Bale} {et~al.}(2016){Bale}, {Goetz}, {Harvey}, {Turin}, {Bonnell},
  {Dudok de Wit}, {Ergun}, {MacDowall}, {Pulupa}, {Andre}, {Bolton},
  {Bougeret}, {Bowen}, {Burgess}, {Cattell}, {Chandran}, {Chaston}, {Chen},
  {Choi}, {Connerney}, {Cranmer}, {Diaz-Aguado}, {Donakowski}, {Drake},
  {Farrell}, {Fergeau}, {Fermin}, {Fischer}, {Fox}, {Glaser}, {Goldstein},
  {Gordon}, {Hanson}, {Harris}, {Hayes}, {Hinze}, {Hollweg}, {Horbury},
  {Howard}, {Hoxie}, {Jannet}, {Karlsson}, {Kasper}, {Kellogg}, {Kien},
  {Klimchuk}, {Krasnoselskikh}, {Krucker}, {Lynch}, {Maksimovic}, {Malaspina},
  {Marker}, {Martin}, {Martinez-Oliveros}, {McCauley}, {McComas}, {McDonald},
  {Meyer-Vernet}, {Moncuquet}, {Monson}, {Mozer}, {Murphy}, {Odom},
  {Oliverson}, {Olson}, {Parker}, {Pankow}, {Phan}, {Quataert}, {Quinn},
  {Ruplin}, {Salem}, {Seitz}, {Sheppard}, {Siy}, {Stevens}, {Summers}, {Szabo},
  {Timofeeva}, {Vaivads}, {Velli}, {Yehle}, {Werthimer}, \&
  {Wygant}}]{2016SSRv..204...49B}
{Bale}, S.~D., {Goetz}, K., {Harvey}, P.~R., {et~al.} 2016, \ssr, 204, 49,
  \dodoi{10.1007/s11214-016-0244-5}

\bibitem[{Bale {et~al.}(2020)Bale, Pulupa, Goetz, Bonnell, Drake, de~Wit,
  Halekas, Harvey, Kasper, MacDowall, Malaspina, \&
  Moncuquet}]{psp_crossref_bale}
Bale, S.~D., Pulupa, M., Goetz, K., {et~al.} 2020, \apj, this issue

\bibitem[{{Bonnin} {et~al.}(2008){Bonnin}, {Hoang}, \&
  {Maksimovic}}]{2008A&A...489..419B}
{Bonnin}, X., {Hoang}, S., \& {Maksimovic}, M. 2008, \aap, 489, 419,
  \dodoi{10.1051/0004-6361:200809777}

\bibitem[{{Bougeret} {et~al.}(1995){Bougeret}, {Kaiser}, {Kellogg}, {Manning},
  {Goetz}, {Monson}, {Monge}, {Friel}, {Meetre}, {Perche}, {Sitruk}, \&
  {Hoang}}]{1995SSRv...71..231B}
{Bougeret}, J.~L., {Kaiser}, M.~L., {Kellogg}, P.~J., {et~al.} 1995, \ssr, 71,
  231, \dodoi{10.1007/BF00751331}

\bibitem[{{Bougeret} {et~al.}(2008){Bougeret}, {Goetz}, {Kaiser}, {Bale},
  {Kellogg}, {Maksimovic}, {Monge}, {Monson}, {Astier}, {Davy}, {Dekkali},
  {Hinze}, {Manning}, {Aguilar-Rodriguez}, {Bonnin}, {Briand}, {Cairns},
  {Cattell}, {Cecconi}, {Eastwood}, {Ergun}, {Fainberg}, {Hoang}, {Huttunen},
  {Krucker}, {Lecacheux}, {MacDowall}, {Macher}, {Mangeney}, {Meetre},
  {Moussas}, {Nguyen}, {Oswald}, {Pulupa}, {Reiner}, {Robinson}, {Rucker},
  {Salem}, {Santolik}, {Silvis}, {Ullrich}, {Zarka}, \&
  {Zouganelis}}]{2008SSRv..136..487B}
{Bougeret}, J.~L., {Goetz}, K., {Kaiser}, M.~L., {et~al.} 2008, \ssr, 136, 487,
  \dodoi{10.1007/s11214-007-9298-8}

\bibitem[{{Cecconi}(2019)}]{2019arXiv190103599C}
{Cecconi}, B. 2019, arXiv e-prints, arXiv:1901.03599.
\newblock \doarXiv{1901.03599}

\bibitem[{{Cecconi} {et~al.}(2008){Cecconi}, {Bonnin}, {Hoang}, {Maksimovic},
  {Bale}, {Bougeret}, {Goetz}, {Lecacheux}, {Reiner}, {Rucker}, \&
  {Zarka}}]{2008SSRv..136..549C}
{Cecconi}, B., {Bonnin}, X., {Hoang}, S., {et~al.} 2008, \ssr, 136, 549,
  \dodoi{10.1007/s11214-007-9255-6}

\bibitem[{{de La Noe} \& {Boischot}(1972)}]{1972A&A....20...55D}
{de La Noe}, J., \& {Boischot}, A. 1972, \aap, 20, 55

\bibitem[{{Dulk} {et~al.}(1998){Dulk}, {Leblanc}, {Robinson}, {Bougeret}, \&
  {Lin}}]{1998JGR...10317223D}
{Dulk}, G.~A., {Leblanc}, Y., {Robinson}, P.~A., {Bougeret}, J.-L., \& {Lin},
  R.~P. 1998, \jgr, 103, 17223, \dodoi{10.1029/97JA03061}

\bibitem[{{Dulk} \& {McLean}(1978)}]{1978SoPh...57..279D}
{Dulk}, G.~A., \& {McLean}, D.~J. 1978, \solphys, 57, 279,
  \dodoi{10.1007/BF00160102}

\bibitem[{{Dulk} \& {Suzuki}(1980)}]{1980A&A....88..203D}
{Dulk}, G.~A., \& {Suzuki}, S. 1980, \aap, 88, 203

\bibitem[{{Dulk} {et~al.}(1984){Dulk}, {Suzuki}, \&
  {Sheridan}}]{1984A&A...130...39D}
{Dulk}, G.~A., {Suzuki}, S., \& {Sheridan}, K.~V. 1984, \aap, 130, 39

\bibitem[{{Eastwood} {et~al.}(2009){Eastwood}, {Bale}, {Maksimovic},
  {Zouganelis}, {Goetz}, {Kaiser}, \& {Bougeret}}]{2009RaSc...44.4012E}
{Eastwood}, J.~P., {Bale}, S.~D., {Maksimovic}, M., {et~al.} 2009, Radio
  Science, 44, RS4012, \dodoi{10.1029/2009RS004146}

\bibitem[{{Eastwood} {et~al.}(2010){Eastwood}, {Wheatland}, {Hudson},
  {Krucker}, {Bale}, {Maksimovic}, {Goetz}, \&
  {Bougeret}}]{2010ApJ...708L..95E}
{Eastwood}, J.~P., {Wheatland}, M.~S., {Hudson}, H.~S., {et~al.} 2010, \apjl,
  708, L95, \dodoi{10.1088/2041-8205/708/2/L95}

\bibitem[{{Ergun} {et~al.}(1998){Ergun}, {Larson}, {Lin}, {McFadden},
  {Carlson}, {Anderson}, {Muschietti}, {McCarthy}, {Parks}, {Reme}, {Bosqued},
  {D'Uston}, {Sanderson}, {Wenzel}, {Kaiser}, {Lepping}, {Bale}, {Kellogg}, \&
  {Bougeret}}]{1998ApJ...503..435E}
{Ergun}, R.~E., {Larson}, D., {Lin}, R.~P., {et~al.} 1998, \apj, 503, 435,
  \dodoi{10.1086/305954}

\bibitem[{{Fainberg} {et~al.}(1985){Fainberg}, {Hoang}, \&
  {Manning}}]{1985A&A...153..145F}
{Fainberg}, J., {Hoang}, S., \& {Manning}, R. 1985, \aap, 153, 145

\bibitem[{{Fitzenreiter} {et~al.}(1976){Fitzenreiter}, {Fainberg}, \&
  {Bundy}}]{1976SoPh...46..465F}
{Fitzenreiter}, R.~J., {Fainberg}, J., \& {Bundy}, R.~B. 1976, \solphys, 46,
  465, \dodoi{10.1007/BF00149870}

\bibitem[{{Fox} {et~al.}(2016){Fox}, {Velli}, {Bale}, {Decker}, {Driesman},
  {Howard}, {Kasper}, {Kinnison}, {Kusterer}, {Lario}, {Lockwood}, {McComas},
  {Raouafi}, \& {Szabo}}]{2016SSRv..204....7F}
{Fox}, N.~J., {Velli}, M.~C., {Bale}, S.~D., {et~al.} 2016, \ssr, 204, 7,
  \dodoi{10.1007/s11214-015-0211-6}

\bibitem[{{Ginzburg} \& {Zhelezniakov}(1958)}]{1958SvA.....2..653G}
{Ginzburg}, V.~L., \& {Zhelezniakov}, V.~V. 1958, \sovast, 2, 653

\bibitem[{{Gurnett} \& {Anderson}(1976)}]{1976Sci...194.1159G}
{Gurnett}, D.~A., \& {Anderson}, R.~R. 1976, Science, 194, 1159,
  \dodoi{10.1126/science.194.4270.1159}

\bibitem[{Halekas {et~al.}(2020)Halekas, Whittlesey, Larson, McGinnis,
  Maksimovic, Berthomier, Kasper, Case, Korreck, Stevens, Klein, Bale,
  MacDowall, Pulupa, Malaspina, Goetz, \& Harvey}]{psp_crossref_halekas}
Halekas, J.~S., Whittlesey, P., Larson, D.~E., {et~al.} 2020, \apj, this issue

\bibitem[{{Hanasz} {et~al.}(1980){Hanasz}, {Schreiber}, \&
  {Aksenov}}]{1980A&A....91..311H}
{Hanasz}, J., {Schreiber}, R., \& {Aksenov}, V.~I. 1980, \aap, 91, 311

\bibitem[{{Heiles} {et~al.}(2001){Heiles}, {Perillat}, {Nolan}, {Lorimer},
  {Bhat}, {Ghosh}, {Lewis}, {O'Neil}, {Salter}, \&
  {Stanimirovic}}]{2001PASP..113.1274H}
{Heiles}, C., {Perillat}, P., {Nolan}, M., {et~al.} 2001, \pasp, 113, 1274,
  \dodoi{10.1086/323289}

\bibitem[{{Kellogg}(1980)}]{1980ApJ...236..696K}
{Kellogg}, P.~J. 1980, \apj, 236, 696, \dodoi{10.1086/157789}

\bibitem[{{Kim} {et~al.}(2007){Kim}, {Cairns}, \&
  {Robinson}}]{2007PhRvL..99a5003K}
{Kim}, E.-H., {Cairns}, I.~H., \& {Robinson}, P.~A. 2007, \prl, 99, 015003,
  \dodoi{10.1103/PhysRevLett.99.015003}

\bibitem[{{Kontar} {et~al.}(2017){Kontar}, {Yu}, {Kuznetsov}, {Emslie},
  {Alcock}, {Jeffrey}, {Melnik}, {Bian}, \&
  {Subramanian}}]{2017NatCo...8.1515K}
{Kontar}, E.~P., {Yu}, S., {Kuznetsov}, A.~A., {et~al.} 2017, Nature
  Communications, 8, 1515, \dodoi{10.1038/s41467-017-01307-8}

\bibitem[{{Krupar} {et~al.}(2014{\natexlab{a}}){Krupar}, {Maksimovic},
  {Santolik}, {Cecconi}, \& {Kruparova}}]{2014SoPh..289.4633K}
{Krupar}, V., {Maksimovic}, M., {Santolik}, O., {Cecconi}, B., \& {Kruparova},
  O. 2014{\natexlab{a}}, \solphys, 289, 4633, \dodoi{10.1007/s11207-014-0601-z}

\bibitem[{{Krupar} {et~al.}(2014{\natexlab{b}}){Krupar}, {Maksimovic},
  {Santolik}, {Kontar}, {Cecconi}, {Hoang}, {Kruparova}, {Soucek}, {Reid}, \&
  {Zaslavsky}}]{2014SoPh..289.3121K}
{Krupar}, V., {Maksimovic}, M., {Santolik}, O., {et~al.} 2014{\natexlab{b}},
  \solphys, 289, 3121, \dodoi{10.1007/s11207-014-0522-x}

\bibitem[{{Krupar} {et~al.}(2018){Krupar}, {Maksimovic}, {Kontar}, {Zaslavsky},
  {Santolik}, {Soucek}, {Kruparova}, {Eastwood}, \&
  {Szabo}}]{2018ApJ...857...82K}
{Krupar}, V., {Maksimovic}, M., {Kontar}, E.~P., {et~al.} 2018, \apj, 857, 82,
  \dodoi{10.3847/1538-4357/aab60f}

\bibitem[{{Leblanc} {et~al.}(1998){Leblanc}, {Dulk}, \&
  {Bougeret}}]{1998SoPh..183..165L}
{Leblanc}, Y., {Dulk}, G.~A., \& {Bougeret}, J.-L. 1998, \solphys, 183, 165,
  \dodoi{10.1023/A:1005049730506}

\bibitem[{{Lecacheux}(2011)}]{2011pre7.conf...13L}
{Lecacheux}, A. 2011, in Planetary, Solar and Heliospheric Radio Emissions (PRE
  VII), ed. H.~O. {Rucker}, W.~S. {Kurth}, P.~{Louarn}, \& G.~{Fischer}, 13--35

\bibitem[{{Lecacheux} {et~al.}(1989){Lecacheux}, {Steinberg}, {Hoang}, \&
  {Dulk}}]{1989A&A...217..237L}
{Lecacheux}, A., {Steinberg}, J.~L., {Hoang}, S., \& {Dulk}, G.~A. 1989, \aap,
  217, 237

\bibitem[{{Lin} {et~al.}(1981){Lin}, {Potter}, {Gurnett}, \&
  {Scarf}}]{1981ApJ...251..364L}
{Lin}, R.~P., {Potter}, D.~W., {Gurnett}, D.~A., \& {Scarf}, F.~L. 1981, \apj,
  251, 364, \dodoi{10.1086/159471}

\bibitem[{Maksimovic {et~al.}(2020)Maksimovic, Bale, Bercic, Bonnell, Case,
  de~Wit, Goetz, Halekas, Harvey, Issautier, Kasper, Korreck, Jagarlamudi,
  Lahmiti, Larson, Lecacheux, Livi, MacDowall, Malaspina, M.~M. Martinovic~and,
  Moncuquet, Pulupa, Stevens, Stverak, Velli, \&
  Whittlesey3}]{psp_crossref_maksimovic}
Maksimovic, M., Bale, S.~D., Bercic, L., {et~al.} 2020, \apj, this issue

\bibitem[{{Malaspina} {et~al.}(2020){Malaspina}, Halekas, Berčič, Larson,
  Whittlesey, Bale, Bonnell, de~Wit, Ergun, Howes, Goetz, Goodrich, Harvey,
  MacDowall, Pulupa, Case, Kasper, Korreck, Livi, \&
  Stevens}]{psp_crossref_malaspina}
{Malaspina}, D.~M., Halekas, J., Berčič, L., {et~al.} 2020, \apj, this
  issue

\bibitem[{{McComas} {et~al.}(2016){McComas}, {Alexander}, {Angold}, {Bale},
  {Beebe}, {Birdwell}, {Boyle}, {Burgum}, {Burnham}, {Christian}, {Cook},
  {Cooper}, {Cummings}, {Davis}, {Desai}, {Dickinson}, {Dirks}, {Do}, {Fox},
  {Giacalone}, {Gold}, {Gurnee}, {Hayes}, {Hill}, {Kasper}, {Kecman}, {Klemic},
  {Krimigis}, {Labrador}, {Layman}, {Leske}, {Livi}, {Matthaeus}, {McNutt},
  {Mewaldt}, {Mitchell}, {Nelson}, {Parker}, {Rankin}, {Roelof}, {Schwadron},
  {Seifert}, {Shuman}, {Stokes}, {Stone}, {Vandegriff}, {Velli}, {von
  Rosenvinge}, {Weidner}, {Wiedenbeck}, \& {Wilson}}]{2016SSRv..204..187M}
{McComas}, D.~J., {Alexander}, N., {Angold}, N., {et~al.} 2016, \ssr, 204, 187,
  \dodoi{10.1007/s11214-014-0059-1}

\bibitem[{{Melrose}(2006)}]{2006ApJ...637.1113M}
{Melrose}, D.~B. 2006, \apj, 637, 1113, \dodoi{10.1086/498499}

\bibitem[{{Melrose}(2017)}]{2017RvMPP...1....5M}
---. 2017, Reviews of Modern Plasma Physics, 1, 5,
  \dodoi{10.1007/s41614-017-0007-0}

\bibitem[{{Meyer-Vernet} \& {Perche}(1989)}]{1989JGR....94.2405M}
{Meyer-Vernet}, N., \& {Perche}, C. 1989, \jgr, 94, 2405,
  \dodoi{10.1029/JA094iA03p02405}

\bibitem[{Moncuquet {et~al.}(2020)Moncuquet, Meyer-Vernet, Issautier, Pulupa,
  Bonnell, Bale, de~Wit, Goetz, Griton, Harvey, MacDowall, Maksimovic, \&
  Malaspina}]{psp_crossref_moncuquet}
Moncuquet, M., Meyer-Vernet, N., Issautier, K., {et~al.} 2020, \apj, this issue

\bibitem[{{Novaco} \& {Brown}(1978)}]{1978ApJ...221..114N}
{Novaco}, J.~C., \& {Brown}, L.~W. 1978, \apj, 221, 114, \dodoi{10.1086/156009}

\bibitem[{{Pulupa} {et~al.}(2017){Pulupa}, {Bale}, {Bonnell}, {Bowen},
  {Carruth}, {Goetz}, {Gordon}, {Harvey}, {Maksimovic},
  {Mart{\'\i}nez-Oliveros}, {Moncuquet}, {Saint-Hilaire}, {Seitz}, \&
  {Sundkvist}}]{2017JGRA..122.2836P}
{Pulupa}, M., {Bale}, S.~D., {Bonnell}, J.~W., {et~al.} 2017, Journal of
  Geophysical Research (Space Physics), 122, 2836, \dodoi{10.1002/2016JA023345}

\bibitem[{{Reid} \& {Ratcliffe}(2014)}]{2014RAA....14..773R}
{Reid}, H. A.~S., \& {Ratcliffe}, H. 2014, Research in Astronomy and
  Astrophysics, 14, 773, \dodoi{10.1088/1674-4527/14/7/003}

\bibitem[{{Reiner} {et~al.}(2007){Reiner}, {Fainberg}, {Kaiser}, \&
  {Bougeret}}]{2007SoPh..241..351R}
{Reiner}, M.~J., {Fainberg}, J., {Kaiser}, M.~L., \& {Bougeret}, J.~L. 2007,
  \solphys, 241, 351, \dodoi{10.1007/s11207-007-0277-8}

\bibitem[{{Reiner} {et~al.}(2009){Reiner}, {Goetz}, {Fainberg}, {Kaiser},
  {Maksimovic}, {Cecconi}, {Hoang}, {Bale}, \&
  {Bougeret}}]{2009SoPh..259..255R}
{Reiner}, M.~J., {Goetz}, K., {Fainberg}, J., {et~al.} 2009, \solphys, 259,
  255, \dodoi{10.1007/s11207-009-9404-z}

\bibitem[{{Sasikumar Raja} \& {Ramesh}(2013)}]{2013ApJ...775...38S}
{Sasikumar Raja}, K., \& {Ramesh}, R. 2013, \apj, 775, 38,
  \dodoi{10.1088/0004-637X/775/1/38}

\bibitem[{{Sharykin} {et~al.}(2018){Sharykin}, {Kontar}, \&
  {Kuznetsov}}]{2018SoPh..293..115S}
{Sharykin}, I.~N., {Kontar}, E.~P., \& {Kuznetsov}, A.~A. 2018, \solphys, 293,
  115, \dodoi{10.1007/s11207-018-1333-2}

\bibitem[{{Stewart}(1985)}]{1985SoPh...96..381S}
{Stewart}, R.~T. 1985, \solphys, 96, 381, \dodoi{10.1007/BF00149692}

\bibitem[{{Suzuki} \& {Dulk}(1985)}]{1985srph.book..289S}
{Suzuki}, S., \& {Dulk}, G.~A. 1985, {Bursts of type III and type V.}, ed.
  D.~J. {McLean} \& N.~R. {Labrum}, 289--332

\bibitem[{{Thejappa} {et~al.}(2012){Thejappa}, {MacDowall}, \&
  {Bergamo}}]{2012ApJ...745..187T}
{Thejappa}, G., {MacDowall}, R.~J., \& {Bergamo}, M. 2012, \apj, 745, 187,
  \dodoi{10.1088/0004-637X/745/2/187}

\bibitem[{{Wheatland}(2004)}]{2004ApJ...609.1134W}
{Wheatland}, M.~S. 2004, \apj, 609, 1134, \dodoi{10.1086/421261}

\bibitem[{{Wild} \& {McCready}(1950)}]{1950AuSRA...3..387W}
{Wild}, J.~P., \& {McCready}, L.~L. 1950, Australian Journal of Scientific
  Research A Physical Sciences, 3, 387, \dodoi{10.1071/PH500387}

\bibitem[{{Zaslavsky} {et~al.}(2011){Zaslavsky}, {Meyer-Vernet}, {Hoang},
  {Maksimovic}, \& {Bale}}]{2011RaSc...46.2008Z}
{Zaslavsky}, A., {Meyer-Vernet}, N., {Hoang}, S., {Maksimovic}, M., \& {Bale},
  S.~D. 2011, Radio Science, 46, RS2008, \dodoi{10.1029/2010RS004464}

\end{thebibliography}


\listofchanges

\end{document}